\documentstyle[11pt]{article}

\newcommand{\be}{\begin{equation}}
\newcommand{\ee}{\end{equation}}
\newcommand{\ba}{\begin{eqnarray}}
\newcommand{\ea}{\end{eqnarray}}

\begin{document}

\title{\vspace{-3.0cm} \hspace{0.0cm} \hspace*{\fill} \\[-5.5ex]
\hspace*{\fill}{\normalsize LA-UR-96-1614} \\[1.5ex]
{\huge {\bf Space-time extensions from space-time
densities and Bose-Einstein correlations}}}
\author{B.R. Schlei\thanks{%
E. Mail: schlei@t2.LANL.gov{\ }}\\
{\it Theoretical Division, Los Alamos National Laboratory,
Los Alamos, NM 87545, USA}
}
\date{May 9, 1996}
\maketitle

\begin{abstract}
Using a (3+1)-dimensional solution of the relativistic
Euler-equations for $Pb+Pb$ at 160 $AGeV$, space-time
extensions of kaon emission zones are calculated from
space-time densities and compared to the inverse
widths of two-kaon Bose-Einstein correlation functions.
The comparison shows a satisfactory agreement and it
is concluded that because of the Gaussian shape of
the kaon correlation functions, the space-time parameters
of the kaon source can be calculated directly from
space-time densities.
In the case of intensity interferometry of identical pions
this simplification is not recommended when applying
Gaussian fits because of the present strong effects of
resonance decays. 
The whole discussion is based on the assumption that
hadron emission in ultra-relativistic heavy-ion collisions
is purely chaotic or that coherence is at least negligible.
\end{abstract}

\hbadness=10000

\vspace{-0.5cm}

\newpage

The knowledge about the lifetime and the spatial
extensions of hadronic sources in ultrarelativistic nucleus-nucleus 
collisions constitutes further information in the search for a 
new state of nuclear matter, the quark-gluon plasma (QGP).
In this effort an experimental technique has been 
developed which is dedicated to the measurement
of Bose-Einstein correlations (BEC) \cite{GGLP}.
BEC functions of identical bosons are quantum-statistical
observables, which in general contain information
about ten quantities  \cite{michael} which characterize a hadron
source: lifetimes, longitudinal and
transverse extensions of the chaotic and the coherent
source, temporal and spatial (longitudinal and transverse)
coherence lengths and the chaoticty.
In the case of pion interferometry it has been argued
that a possible coherent subcomponent of the hadron
source is almost inobservable \cite{bernd3} due to the apparently 
large contributions from resonance decays to BEC.
Many parametrizations
of experimental BEC functions are based on a neglect
of a coherent component, resulting in the
description of the hadron emitting source with only three
quantities ({\it e.g.}, lifetime, longitudinal and transverse radii).
In ref. \cite{chapman} it was shown, that in fact even in
the case of a purely chaotic hadron source three quantities
are not adequate to fully parametrize the system under consideration.
{\it E.g.}, a cylindrically symmetric expanding fireball has in general 
to be described with an additional ``cross-term'' radius.
Depending on the concepts of introducing a radius or a lifetime 
into a model of relativistically expanding hadron sources,
it appears to be very likely that a direct relation between interferometry
radii (better: inverse widths of BEC functions) and the true 
space-time characteristics of the source can be established.\\

It is the purpose of this paper to study whether and how the true
space-time features of hadron emitting sources are reflected in
two-particle Bose-Einstein correlation functions of purely chaotic
sources by applying a relativistic hydrodynamical description.
Many hydrodynamical models ({\it cf.} \cite{bernd8} and refs. therein) 
are available which describe the dynamics of relativistic heavy-ion 
collisions. HYLANDER \cite{udo} belongs to the class of models which apply 
(3+1)-dimensional relativistic one-fluid-dynamics. 
It provides fully three-dimensional solutions of the hydrodynamical 
relativistic Euler-equations \cite{euler}. The model has been successfully 
applied to several different heavy ion collisions at SPS energies 
\cite{marburg}. Here some further results of the HYLANDER 
analysis \cite{bernd8,bernd7} of $Pb+Pb$ at 160 $AGeV$ (central
collisions) will be presented.\\

From the hydrodynamical analysis \cite{bernd8,bernd7} of $Pb+Pb$ at 
160 $AGeV$, it has been found that the $K^-$ single inclusive spectra 
consist mainly of directly emitted kaons; less than $10\%$ come from the 
$K^\star$ resonance.
Because the $K^\star$ resonance contributions do not exhibit any significant 
effect in the inverse widths and the functional shapes
of calculated $K^-K^-$ BEC functions, 
these resonance contributions have been neglected. 
Introducing the average and the relative four-momentum, 
$K^\nu \equiv \textstyle{\frac{1}{2}}(k_1^\nu + k_2^\nu)$ and 
$q^\nu \equiv k_1^\nu - k_2^\nu$, of two identical kaons emitted with their
individual four-momenta, $k_1^\nu$ and $k_2^\nu$, respectively,  
the Bose-Einstein correlation function of the kaon-pair is in terms of 
relativistic hydrodynamics defined as ({\it cf.} refs. 
\cite{bernd3,bernd1,bernd2})
\begin{equation}
C_2(\vec{k}_1,\vec{k}_2)\:=\:1\:+\:
\frac{| \int d^4 x\ g^{dir}_K (x_\nu,K^\nu)\:  
\displaystyle{e^{i q^\nu x_\nu}}\: |^2}
{\int d^4 x\ g^{dir}_K (x_\nu,k_1^\nu) \cdot 
\int d^4 x\ g^{dir}_K (x_\nu,k_2^\nu)} \:,
\label{eq:C2def}
\end{equation}

where $q^0=E_1-E_2$, $K^0=(E_1+E_2)/2$, and

\begin{equation}
g^{dir}_K (x_\nu,k^\nu)\:=\: \frac{1}{(2\pi)^3}
\int_\Sigma \frac{k^\nu d\sigma_\nu(x^\prime_\nu) \:
\delta^4(x_\nu - x^\prime_\nu)}
{\displaystyle{\exp\left[\frac{
k^\nu  u_\nu(x^\prime_\nu)-S \mu_S(x^\prime_\nu)}
{T_f(x^\prime_\nu)}\right]}-1}\:,\quad x^\prime_\nu \in \Sigma\:.
\label{eq:gdirk}
\end{equation}

In eq. (\ref{eq:gdirk}) $d\sigma_\nu$, $u_\nu$ and $T_f$
are the differential volume element of the freeze-out hypersurface $\Sigma$,
the 4-velocity of the fluid and the freeze-out temperature
at space-time point $x_\nu$, respectively. The quantity $\mu_S$ is 
the strangeness chemical potential at space-time point $x_\nu$.
(For a calculation of BEC of identical pions including the decay of
resonances, {\it cf.} refs. \cite{bernd3,bernd2}.)
The function $g_K^{dir}$ can be interpreted as the quantum analogue of
the mean number of kaons of four-momentum $k^\nu$ at the space-time
point $x_\nu$, {\it i.e.}, one can consider $g_K^{dir}$ as the 
quantum analogue of the space-time density distribution of the source 
emitting kaons of fixed four-momenta $k^\nu$. For the following, let us 
keep the latter interpretation in mind.\\

Fig. 1 shows a comparison of BEC functions of $\pi^-$ (solid lines)
and of $K^-$ (dashed lines) at $K_\perp=0$ and $K_\perp=1$ $GeV/c$ ,
respectively. The dotted lines correspond to the BEC
functions of directly (thermally) produced $\pi^-$.
It can be seen that BEC functions of $\pi^-$ are strongly distorted
by the decay of resonances. In particular, the functional shape
of the BEC functions of $\pi^-$ (solid lines) depends strongly
on the momentum of the pair, {\it i.e.}, it varies between the extremes of
an exponential and a Gaussian form. On the contrary, the functional shape
of the BEC functions of thermally produced $\pi^-$ and those of $K^-$
is of almost perfect Gaussian shape, independent of the average 
momentum of the pair.
In case of the shown transverse momentum-dependent BEC functions, the
ones for thermal pions and kaons are almost identical. 
In the following we shall discuss only BEC functions of $K^-$ pairs.\\

Let us now require for a fit to a two-particle BEC function: 
\begin{description}
\item[(a)] The selected functional shape of the fit should reproduce the  
entire two-particle BEC function.
\item[(b)] The selected fitting function should respond to the changes
in the average momentum of the pair.
\end{description}

These demands ensure that a reconstruction of the original
BEC function is possible through the knowledge of only the 
fitted parameters.
In the case of kaon interferometry, (a) and (b) are easily to
fulfill. The BEC functions can be fitted with a Gaussian,
because their functional shape is independent of the choice of the
average momentum of the kaon-pair under consideration.
Therefore, we fit our results to (for the choice of the variables, 
{\it cf., e.g.} refs. \cite{chapman,bertsch})
\begin{equation}
C_2(\vec{k}_1,\vec{k}_2)\:=\: 1\:+\:\exp
\left[-\:q_\parallel^2 R_{long}^2(\vec{K}) \:-\:q_{side}^2
R_{side}^2(\vec{K})\:-\:q_{out}^2 R_{out}^2(\vec{K})\:-2\:q_\parallel 
q_{out} R_{cross}^2(\vec{K})\right] \:.
\label{eq:fit}
\end{equation} 

A characteristic property of particle production from an expanding source
is a correlation between the space-time point where a particle is emitted
and its energy-momentum \cite{pratt}. As a consequence, the inverse widths
$R_i(\vec{K})$ ($i={\parallel}, {side}, {out}, {cross}$) extracted from 
Bose-Einstein correlation functions show a characteristic dependence of the 
average momentum of the pair, $\vec{K}$ ({\it cf.} also Fig. 1).\\

Figs. 2a and 2b show the calculations for $R_{side}$, $R_{out}$ and $R_{long}$
as functions of rapidity $y_K$ and transverse momentum $K_\perp$ 
from the fit of eq. (\ref{eq:fit}) to the BEC functions in one 
$q$-dimension, {\it i.e.}, the remaining momentum differences have been 
set equal to zero. $R_{cross}$ has therefore not been extracted from a fit. 
For the specific choices of $y_K=0$ and $K_\perp=0$, the inverse widths of 
the kaon BEC from $Pb+Pb$ at 160 $AGeV$ have already been shown\footnote{
Note the reduction of the inverse widths by a factor $1/\sqrt{2}$. 
The inverse widths of ref. \cite{bernd7} have been obtained with a different
parametrization.} in ref. \cite{bernd7}.
Since measurements of kaon interferometry radii have not been published
yet for this particular heavy-ion reaction, the results shown in Figs. 2a 
and 2b have to be considered as further predictions of the hydrodynamical 
analysis presented in refs. \cite{bernd8,bernd7}.\\

In the following, we introduce an approximation \cite{csorgo1,chapman2} 
of the original
definition ({\it cf.} eq. (\ref{eq:C2def})) of the two-particle BEC,
\begin{equation}
C_2(\vec{k}_1,\vec{k}_2)\:\approx\:1\:+\:
\frac{| \int d^4 x\ g^{dir}_K (x_\nu,K^\nu)\:  
\displaystyle{e^{i q^\nu  x_\nu}}\: |^2}
{| \int d^4 x\ g^{dir}_K (x_\nu,K^\nu) \: |^2} 
\:\equiv\: \tilde{\cal C}_2(\vec{K},\vec{q}\:)\:,
\label{eq:C2app}
\end{equation}

where $q^0=E_1-E_2$ and $E_K \equiv K^0=\sqrt{m^2+\vec{K}^2}$. In connection
with analytical parametrizations of the momentum-dependent space-time
density distribution $g_K^{dir}$, the approximation eq. (\ref{eq:C2app}) 
has been 
shown to be acceptable \cite{chapman,chapman2,wiedemann,csorgo2}. 
Here, we are going to apply the hydrodynamical 
solution already presented in refs. \cite{bernd8,bernd7} for $g_K^{dir}$ 
according to eq. (\ref{eq:gdirk}).\\

Using $x^\nu\equiv (t,x,y,z)$, we obtain for kaon pairs with 
$|\vec{q}\:| \ll E_K$ approximately
\begin{equation}
q^\nu x_\nu\:\approx\:(\beta_\perp q_{out}\:+\:\beta_\parallel q_\parallel)
\:t \:-\: q_{out}\:x\:-\: q_{side}\:y\:-\: q_\parallel \:z \:,
\label{eq:qxapp}
\end{equation}

where $\beta_i=K_i/E_K$ ($i=\:\parallel, \perp$).
For any cylindrically symmetric emission function which
can be roughly expressed in Gaussian form one finds with eq. 
(\ref{eq:qxapp}) when expanding $\exp [i q^\nu  x_\nu ]$ 
in eq. (\ref{eq:C2app}) for $q^\nu x_\nu \ll 1$ the expression 
\begin{eqnarray}
\tilde{\cal C}_2(\vec{K},\vec{q}\:)\:\approx\:1\:+\:\{ 
1&-&q_{side}^2 \langle y^2 \rangle\:
-\: \langle [q_{out}(\beta_\perp t\:-\:x)\:+\:q_\parallel(\beta_\parallel t
\:-\:z)]^2\rangle
\nonumber\\
&+&\langle q_{out}(\beta_\perp t\:-\:x) \:+\:q_\parallel(\beta_\parallel t
\:-\:z)\rangle^2\:+\:{\cal O}[(q^\nu x_\nu)^4]\}\:,
\label{eq:c2tay}
\end{eqnarray}

where
\begin{equation}
\langle \xi \rangle\:\equiv\:\langle \xi \rangle(k^\nu)\:
=\:\frac{\int d^4 x\ \xi\:g^{dir}_K (x_\nu,k^\nu)}
{\int d^4 x\ g^{dir}_K (x_\nu,k^\nu)}\:.
\label{eq:avdens}
\end{equation}

The average $\langle \xi \rangle$ can be considered as an expectation value 
from space-time densities $g_K^{dir}$.
After exponentiating eq. (\ref{eq:c2tay}), 
for any cylindrically symmetric system the BEC function $\tilde{\cal C}_2$
can be expressed for small momentum differences ($q_i R_i \ll 1$) 
in the form of eq. (\ref{eq:fit}). When comparing the coefficients
of the $q_i q_j$ terms, the functions $R_i^2(\vec{K}\:)$ are found to be
given through ({\it cf.} \cite{wiedemann} and refs. therein)
\begin{eqnarray}
R_{side}^2(\vec{K})&\approx&\langle y^2 \rangle
\:=\:\langle y^2 \rangle\:-\:\langle y \rangle^2\:\equiv\:\sigma^2_y\:,
\nonumber\\
R_{out}^2(\vec{K})&\approx&\langle (x\:-\:\beta_\perp t)^2\rangle\:-\:
\langle x\:-\:\beta_\perp t\rangle^2\:\equiv\:
\sigma^2_{x\:-\:\beta_\perp t}\:,
\nonumber\\
R_{long}^2(\vec{K})&\approx&\langle (z\:-\:\beta_\parallel t)^2\rangle\:-\:
\langle z\:-\:\beta_\parallel t\rangle^2\:\equiv
\:\sigma^2_{z\:-\:\beta_\parallel t}\:,
\nonumber\\
R_{cross}^2(\vec{K})&\approx&\langle (x\:-\:\beta_\perp t)(z\:-\:
\beta_\parallel t) 
\rangle\:-\:\langle x\:-\:\beta_\perp t\rangle \langle z\:-
\:\beta_\parallel t\rangle \:\equiv\: 
\sigma_{\:x\:-\:\beta_\perp t,\:z\:-\:\beta_\parallel t}\:.
\label{eq:radii}
\end{eqnarray}

Thus the inverse widths of the two-kaon BEC function can be
calculated directly from the ten space-time averages $\langle t \rangle$,
$\langle x \rangle$, $\langle z \rangle$, $\langle t^2 \rangle$,
$\langle x^2 \rangle$, $\langle y^2 \rangle$, $\langle z^2 \rangle$,
$\langle xt \rangle$, $\langle zt \rangle$ and $\langle xz \rangle$
(note that $\langle y \rangle\equiv 0$ for a cylindrically
symmetric expanding system).  
The quantities $\sigma^2_i$ and $\sigma_{i,j}$ in 
eqs. (\ref{eq:radii}) can be interpreted as the variance ({\it cf.} also ref.
\cite{heiselberg}) of the random variable $i$ and the covariance of the 
random variables $i$ and $j$, respectively (of course, 
$x^\nu = (t, x, y, z) \in \Sigma$).
The inverse widths of kaon BEC functions have therefore a 
geometrical interpretation involving the relativistic kinematics of the
emitted particle.\\

In Figs. 2a and 2b the space-time averages from space-time
densities have been calculated from the r.h.s of eqs. (\ref{eq:radii}) 
and are compared to the inverse widths of kaon BEC functions
obtained from the direct fit (\ref{eq:fit}). The agreement is remarkable,
although $R_{side}$ and $R_{out}$ from fits to kaon BEC functions are always 
slightly underestimated, while $R_{long}$ is always slightly overestimated.
The differences of $R_{long}$, $R_{side}$ and $R_{out}$ of both calculations
have their origin in the approximations 
\mbox{(\ref{eq:C2app}) - (\ref{eq:c2tay})} and in the fact
that the kaon BEC functions are not of perfect Gaussian shape for all average
momenta of the kaon pairs.
The maximal differences are 0.54 $fm$, 0.15 $fm$ and 0.17 $fm$,
when comparing $R_{long}$, $R_{side}$ and $R_{out}$
of both calculations, respectively.
If one is willing to accept errors of the magnitude here presented, 
one is tempted to avoid the calculation and subsequent fit of
two-kaon BEC functions, since the inverse widths $R_i(\vec{K})$ can be 
calculated in a much more effortless way directly from space-time averages.
The feature of BEC of identical kaons to be of Gaussian shape
leads through the knowledge of the functions $R_i(\vec{K})$ always to an
acceptable reconstruction of single BEC functions.
In addition to the inverse widths which have been extracted from
BEC functions of kaons, the cross-term radius $R_{cross}(\vec{K})$ has 
been calculated from the space-time expectation values ({\it cf.} Fig. 2b).\\

Let us now return to the BEC functions of identical pions.
We have found that the inverse widths of BEC of direct pions can be 
also treated as in the case of interferometry of identical kaons,
{\it i.e.}, through the use of effective Gaussians.
The maximal differences when comparing $R_{long}$, $R_{side}$ and $R_{out}$
of both calculations are 0.78 $fm$, 0.20 $fm$ and 0.20 $fm$, respectively.
But it is obvious from the above presentation that an effective Gaussian
would not constitute an appropriate fit for a single BEC function
of all pions, {\it i.e.}, including those from the decay of resonances, because
of the drastic change in the functional shape ({\it cf.} Fig. 1).
When it comes to the fit of experimentally observed BEC of pions,
the situation is even more complicated, although measured BEC might
have a Gaussian shape\footnote{
It is therefore unclear, how the conclusions of ref. \cite{heiselberg}
are affected.}. Experimentally obtained BEC functions
have to be calculated from separate integrations of the numerator
and the denominator of the correlators of BEC functions of single
momentum pairs with respect to detector acceptances \cite{bernd8,future}.\\

To summarize, Bose-Einstein correlation functions of identical kaons 
have been calculated for $Pb+Pb$ at 160 $AGeV$ from a 
(3+1)-dimensional solution of the relativistic Euler-equations.
From the hydrodynamical treatment it has been found that almost all kaon 
BEC are of Gaussian shape, irrespective of the average momenta of the kaon 
pairs under consideration.
Since in the calculations the kaon source was assumed to be completely
chaotic, the BEC of kaons have been parametrized with a Gaussian
and inverse widths of the BEC functions have been extracted.\\
Interpreting the source function $g_K^{dir}(x_\nu,k^\nu)$ as
the quantum analogue to the space-time density distribution of
emitted kaons of fixed four-momentum $k^\nu$, space-time averages of
space-time coordinates of the source have been calculated and are
related to the inverse widths of BEC of kaons.
The comparison of inverse widths of kaon BEC functions
and space-time extensions 
from space-time densities shows a satisfactory agreement. It
is concluded that because of the Gaussian shape of
the kaon correlation functions, the space-time parameters
of the kaon source can be calculated directly from
space-time densities within errors of several 0.1 $fm$.\\
In the case of intensity interferometry of identical pions
this simplification is not recommended when applying
Gaussian fits because of the presence of resonance decays 
the knowledge of only the inverse widths of the two-pion BEC is not sufficient 
to reconstruct the entire correlation function.\\*[1.5em] 

B.R.S. is indebted to U. Heinz for discussions that initiated this work. 
Many helpful discussions with D. Strottman and his kind hospitality 
extended to the author at the Los Alamos National Laboratory are gratefully 
acknowledged. This work has been supported by the University of California 
and the Deutsche Forschunggemeinschaft (DFG). The author acknowledges a DFG 
postdoctoral fellowship.

\newpage
\noindent
{\Large {\bf Figure Captions}}\\

\begin{description}

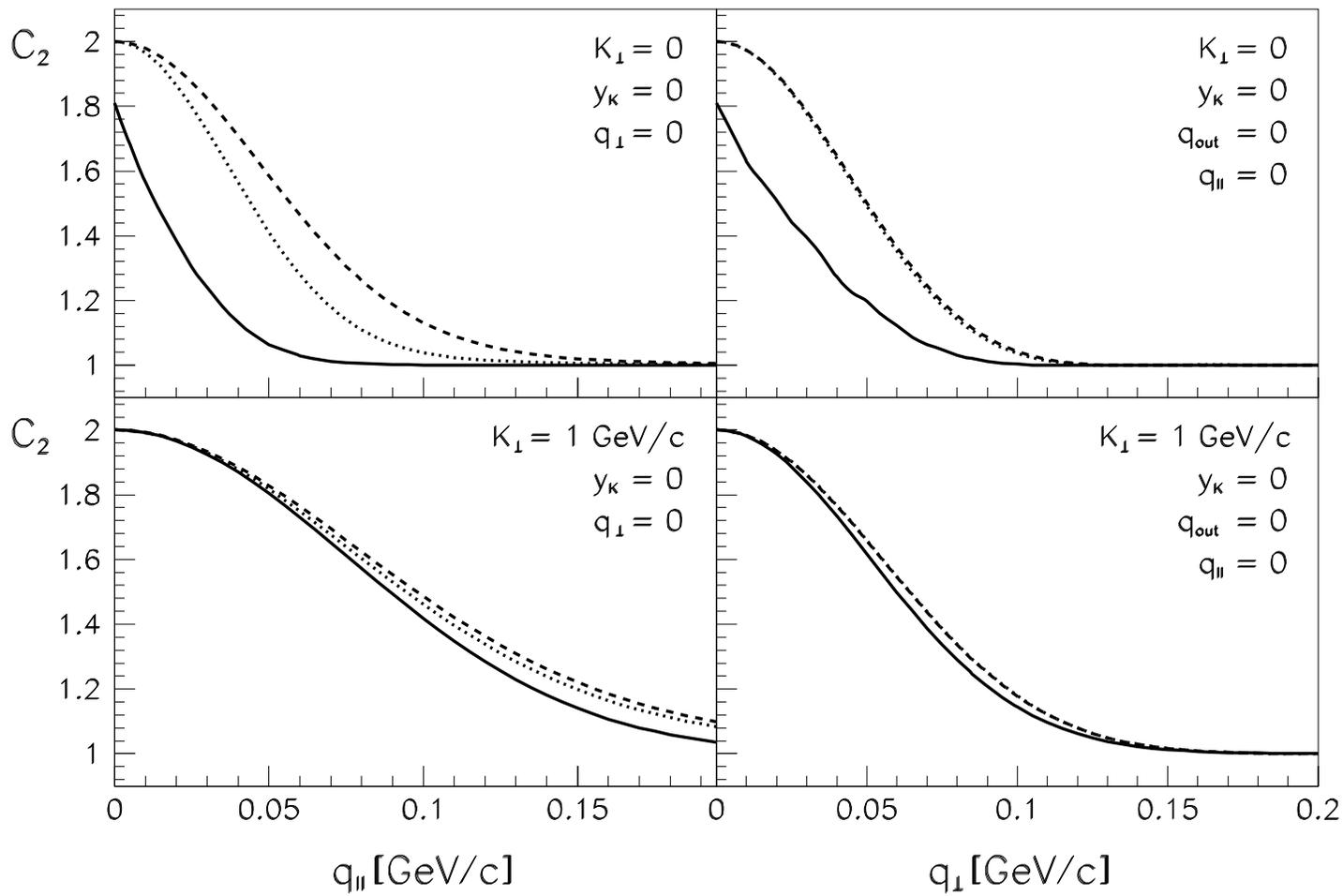
\item[Fig. 1] Examples of Bose-Einstein correlation functions of all 
$\pi^-$ (solid lines), thermal $\pi^-$, and $K^-$ (dashed lines),
for $K_\perp =0$ and $K_\perp =1GeV/c$. The BEC functions have been 
calculated for $Pb+Pb$ at 160 $AGeV$.

\item[Fig. 2a] Momentum-dependence of $R_{side}$ and $R_{out}$ extracted
from Bose-Einstein correlation functions of kaons compared
to the directly from space-time densities calculated ones.

\item[Fig. 2b] Momentum-dependence of $R_{long}$ extracted
from Bose-Einstein correlation functions of kaons (BEC) compared
to the directly from space-time densities calculated ones.
The radius $R_{cross}$ has been calculated from space-time densities
only.

\end{description}

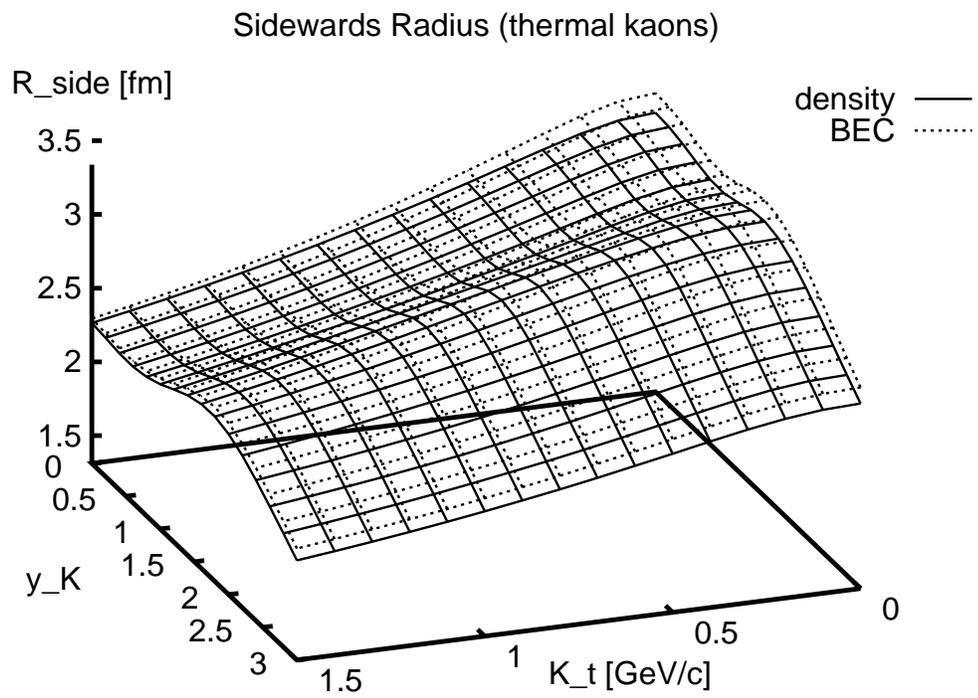
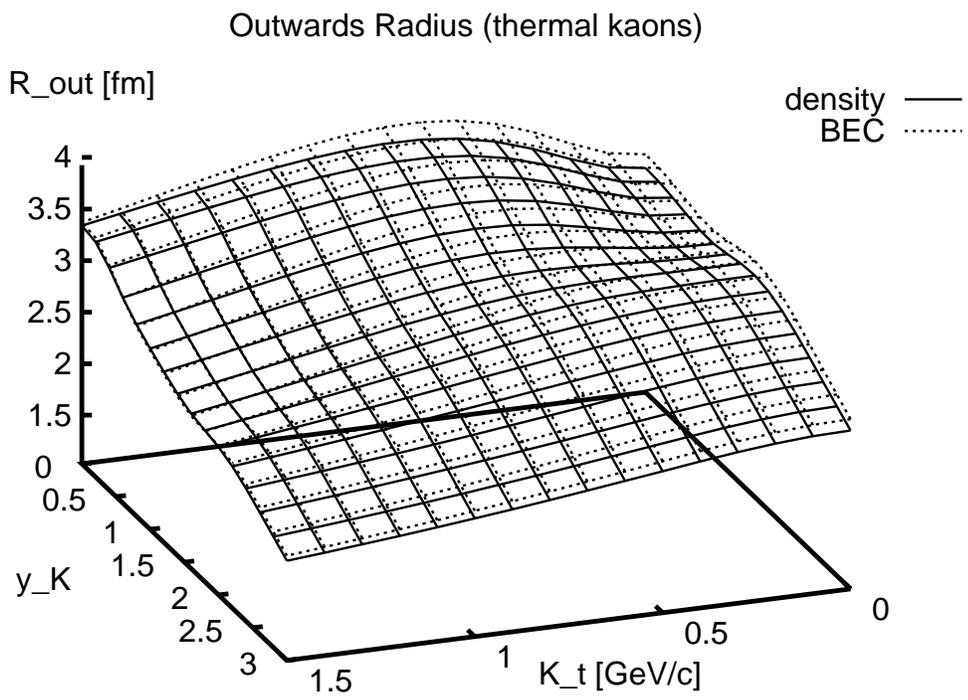
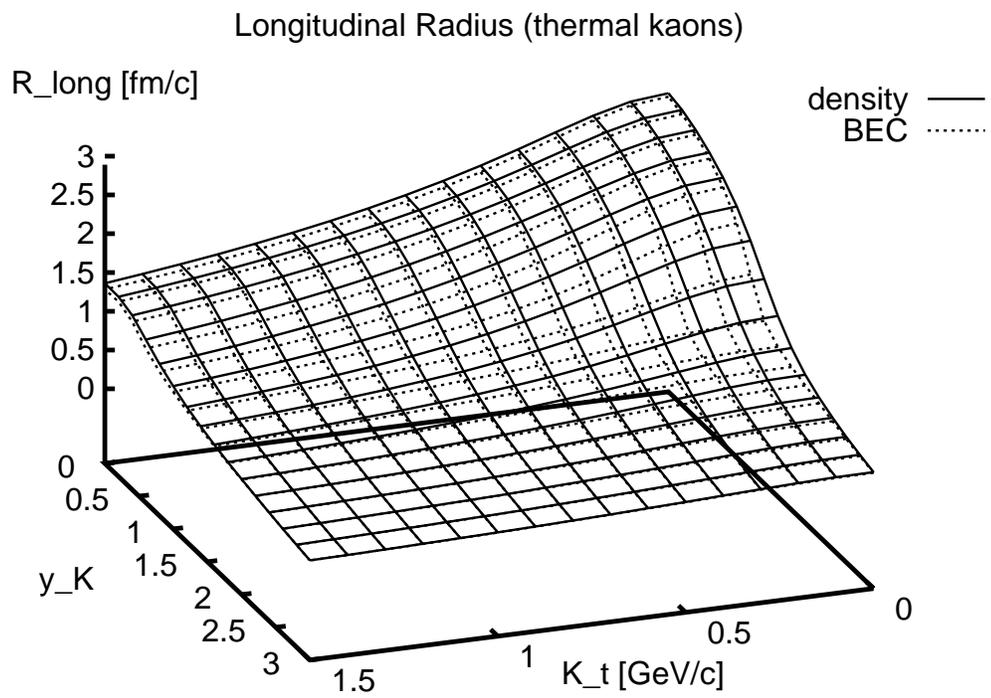
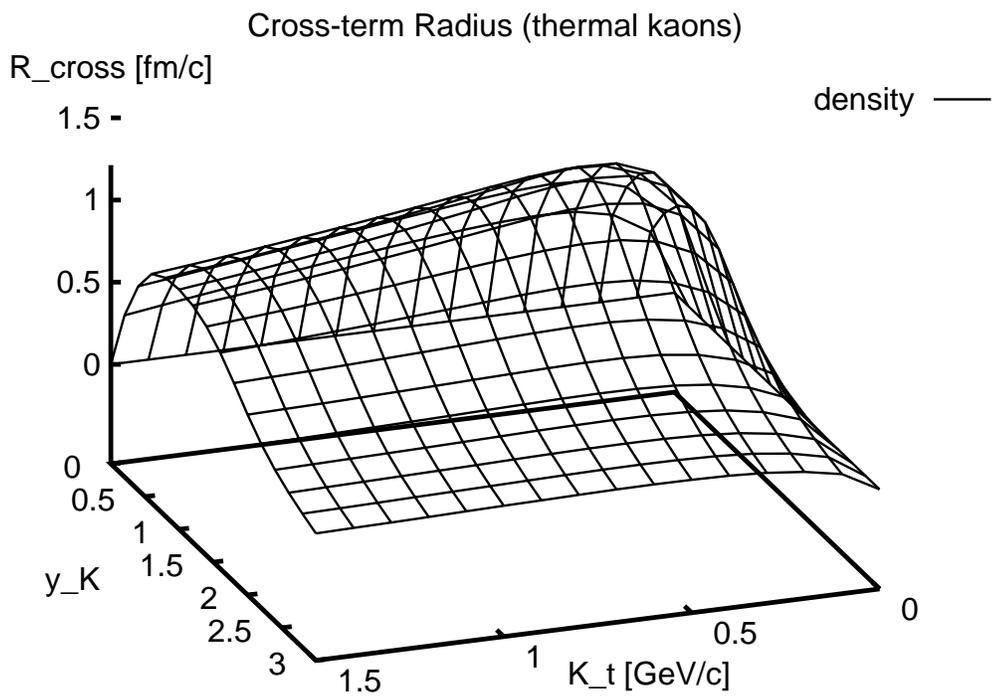
\end{document}